\documentclass[twocolumn,aps,showpacs,amsmath,amssymb,floatfix]{revtex4}
\usepackage{graphics,color}

\begin{document}

\title{
  Optical conductivity of wet DNA
}

\author{
  A. H\"{u}bsch$^{1}$, R. G. Endres$^{2}$, D. L. Cox$^{1,3,4}$, and 
  R. R. P. Singh$^{1,3}$}

  \affiliation{
    $^{1}$Department of Physics, University of California, Davis, CA 95616\\
    $^{2}$NEC Laboratories America, Inc., Princeton, NJ 08540\\
    $^{3}$Center for Biophotonics Science and Technology, University of
    California, Davis, CA 95616\\
    $^{4}$Center for Theoretical Biological Physics, University of California,
    San Diego, CA 92119
  }

\date{\today}

\begin{abstract}
Motivated by recent experiments
we study the optical conductivity of DNA in its natural
environment containing water molecules and counter ions. Our density
functional theory calculations (using SIESTA) for four base pair B-DNA with
order 250 surrounding water molecules suggest a thermally activated doping of
the DNA by water states which generically leads to an electronic contribution
to low-frequency absorption. The main contributions to the doping result from
water near  DNA ends, breaks, or nicks and are thus potentially associated
with temporal or structural defects in the DNA.  
\end{abstract}


\maketitle

The electronic properties of DNA have received considerable scientific
attention in the last few years, motivated both by possible use in molecular
electronics \cite{Ventra} and by speculation that electronic
processes can speed up the
location (and possibly repair) of damage sites in living
cells \cite{BartonElDamage}. A number of experiments have studied
conductance of DNA in different contexts (for a recent review see
Ref.~\onlinecite{Endres_RMP}). Of particular note have been observations of
activated behavior with small gaps, of order 0.1-0.3
eV \cite{KawaiPRL,Helgren},  which stand in 
contrast to the peak optical absorption of DNA in the ultraviolet at 3.5-4 eV.
Evidently, such small activation energies, if not extrinsically
induced, require an intrinsic doping mechanism.

To address this issue, we study several DNA tetramers surrounded by
waters and counterions  using a combination of classical molecular dynamics
(MD) with density functional theory (DFT) employing the local basis set SIESTA
code \cite{SIESTA}. To our knowledge, this is the first time that the
electronic conductivity of DNA has been computed with a full quantum mechanical
treatment of water and counterions  in a dynamically fluctuating
environment, although similar methods have been used before to study charge 
migration in DNA \cite{italianpaper,Landman} {\it without} explicit quantum
treatment of the environment. Our tetramers are in the  fully hydrated
biological form (B-DNA), with order 250 
water molecules.  We find evidence that waters in contact with DNA bases can
dope the DNA with excitation gaps as small as 0.1-0.3 eV. 
We find absorption at low frequencies in agreement with
optical experiments \cite{Helgren} if we increase the number of bases exposed
to water.  
In a cellular environment, these will be associated with temporal or
structural defects, but in many solid-state experiments can also arise from
the flattening and unwinding of the DNA helix.

We also study the peak absorption in DNA in the 
ultraviolet at 3.5-4 eV. 
The calculated frequency dependent
conductivity agrees well with the measured peak
absorption for $\lambda$-phage DNA\cite{Helgren}. While the
peak location is found to be stable, there are distinct features near 
the peak, which
vary with DNA sequence and different environmental configurations
accessed during a typical MD run.  Within the time scales of
the MD (out to 1 ns), our 
calculations are unable to uniquely determine sequence specificity, but some
specificity/fingerprinting may be possible with signal averaging. 

These results confirm that while the DNA is a large-gap material,
environmental conditions can induce states in the gap,
which can in turn lead to a small but non-zero sub-gap absorption.
Such states are also crucial for understanding room temperature thermally 
activated electron dynamics in the DNA, a mechanism that has gained wide
acceptance for observed long-range electron transfer in DNA\cite{Ventra}. 

Electronic structure calculations have previously been
performed for dry A-DNA \cite{Pablo} and for crystallized Z-DNA \cite{Gervasio}
where the effects of solvent and counterions were also addressed. Here,
we extend our recent work \cite{Endres_CM} by considering dynamical
fluctuations of the environment.  Other efforts have been made to study
electronic spectra of DNA in fluctuating environments both
without \cite{Lewis2002} and with \cite{italianpaper,Landman} counterions and
water.  Our work differs from these latter efforts by the explicit quantum
mechanical treatment of water and counterion states.  

To calculate the electronic spectra and optical conductivity we have used the
fully ab initio DFT 
code SIESTA \cite{SIESTA}. It uses Troullier-Martins norm-conserving pseudo
potentials \cite{Troullier} in the Kleinman-Bylander form \cite{Kleinman}. We
have used  the generalized gradient approximation (GGA) for the
exchange-correlation 
energy functional in the version of Perdew, Burke, and Ernzerhof
\cite{Perdew}. SIESTA uses a basis set of numerical atomic orbitals where the
method by 
Sankey and Niklewski \cite{Sankey} is employed. For the DNA we have used a
double-$\zeta$ basis set except for phosphorus 
and the counterions for which the polarization orbitals are also
included. For the surrounding water molecules (about 250) we have
only used the minimal, single-$\zeta$ basis set.

The optical conductivity is given by the Kubo formula 
\begin{eqnarray*}
    \sigma_{1,\mu}(\omega)
    &=&
    \pi \frac{e^{2}}{V}
    \frac{1 - e^{-\beta\hbar\omega}}{\omega}\\
    && \times
    \sum_{n,m}
    \frac{ e^{-\beta\varepsilon_{n}} }{Z} \,
    \delta\left(
      \varepsilon_{n} - \varepsilon_{m} - \hbar\omega
    \right)
    \left|
      \left\langle n \right| j_{\mu} \left| m \right\rangle
    \right|^{2}
\end{eqnarray*}
and depends on the frequency $\omega$. Here, $\varepsilon_{n}$,
$\varepsilon_{m}$ and $|n\rangle$, $|m\rangle$ are the energies and wave
functions of the DFT Hamiltonian, respectively. $V$ denotes the unit cell
volume, and $j_{\mu}$ with $\mu = \perp, \parallel$ is the component of the
current operator perpendicular or parallel to the DNA strand direction.
The conductivity computed here does not include time dependent DFT or scissors
corrections.  

The solvated DNA structures were obtained from 
classical MD simulations with the AMBER7 package \cite{amber7} 
(parm98 force field). The four base-pair long
DNA structures (B-DNA 5'-GAAT-3', 5'-GGGG-3', 5'-AAAA-3', and TT-dimer) 
were initially charge-neutralized with
counterions (either by 6 Na$^+$ or 3 Mg$^{2+}$)
and solvated with about 600 TIP3P water molecules.
A simulation box of dimensions $38 \times 35.4 \times 25 \AA^3$
(z-axis parallel to DNA axis) 
with periodic boundary conditions was applied.
After equilibration at room temperature for 1 ns, 
the trajectory was recorded every 10 ps over a 
simulation time of 1 ns. The resulting 100 snapshots were used
for subsequent analysis. All simulations employed the 
SHAKE method to fix hydrogen-heavy atom distances allowing
a 2 fs integration time step. The cut-off for long-range interactions
was set to 10~$\AA$. 
TT-dimer (segment 5'-ATTA-3' from PDB code 1SM5) 
was further constrained using the BELLY option to preserve 
the distorted DNA structure.

To use the structures from MD simulations in DFT
calculations we had to reduce the number of atoms. We only
kept the first and second solvation shell (approximately 30 water molecules
per nucleotide) so that all water molecules within a 4.7~\AA~radius
were included. We incorporated about 1000
atoms in our DFT calculation so that the computations were quite
time consuming. In particular, the evaluation of the optical conductivity was
very expensive. To speed up the computations we restricted the number of
included unoccupied states which only affect the high-energy spectra above 5
eV. The calculation of optical conductivity for a
given structure snapshot took about 5 days on AMD's Opteron CPU where
approximately 2~GB RAM was needed. All calculations were done
at room temperature. 

\begin{figure}[t]
\begin{center}
  \scalebox{0.5}{
    \includegraphics*[120,290][495,710]{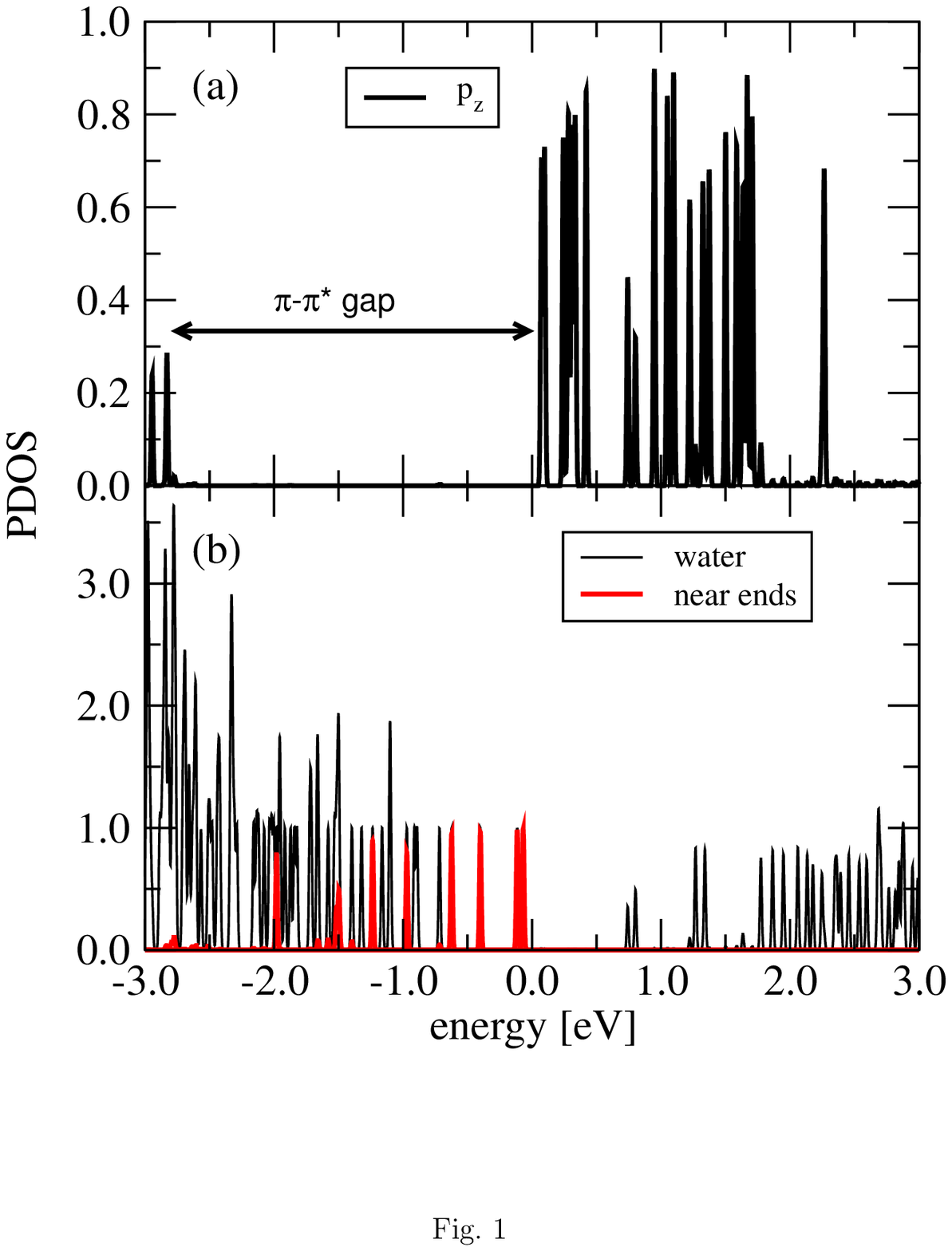}
  }
\end{center}
\caption{
  Typical PDOS of a structure snapshot of a 5'-GAAT-3' sequence with Mg
  counterions. Panel (a) shows the $p_z$ orbitals of atoms in base
  pairs. Panel (b) contains the PDOS of all water molecules (thin black line)
  where the contributions of water molecules near the sequence ends are
  plotted by a thick red line. The Fermi energy is set to $0.0$ eV. 
}
\label{Fig_pdos}
\end{figure}

\begin{figure}[t]
\begin{center}
  \scalebox{0.5}{
    \includegraphics*[75,445][525,705]{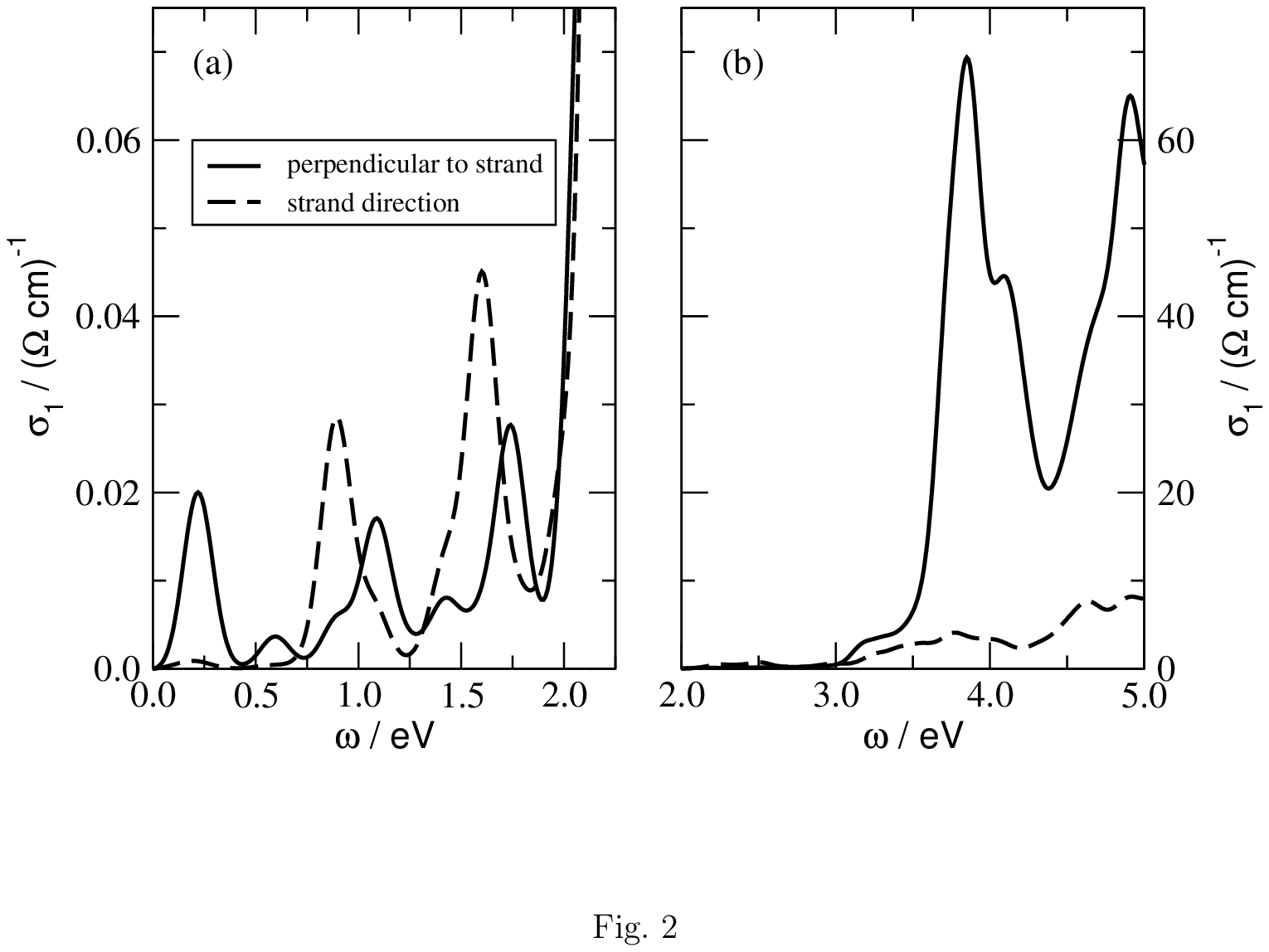}
  }
\end{center}
\caption{
  Polarization dependence of the optical conductivity for a structure snapshot
  of a 5'-GAAT-3' sequence with Na counterions where $\sigma_{1,\perp}(\omega)$
  [$\sigma_{1,\parallel}(\omega)$] is plotted with solid [dashed] lines. 
  Panel (a) shows the low-energy features whereas panel (b) presents
  the range of the $\pi$-$\pi^{*}$ transition. The theoretical
  line spectra are broadened with Gaussian functions of width $0.1$ eV.
}
\label{Fig_pol}
\end{figure}

\begin{figure}[t]
\begin{center}
  \scalebox{0.5}{
    \includegraphics*[75,415][535,705]{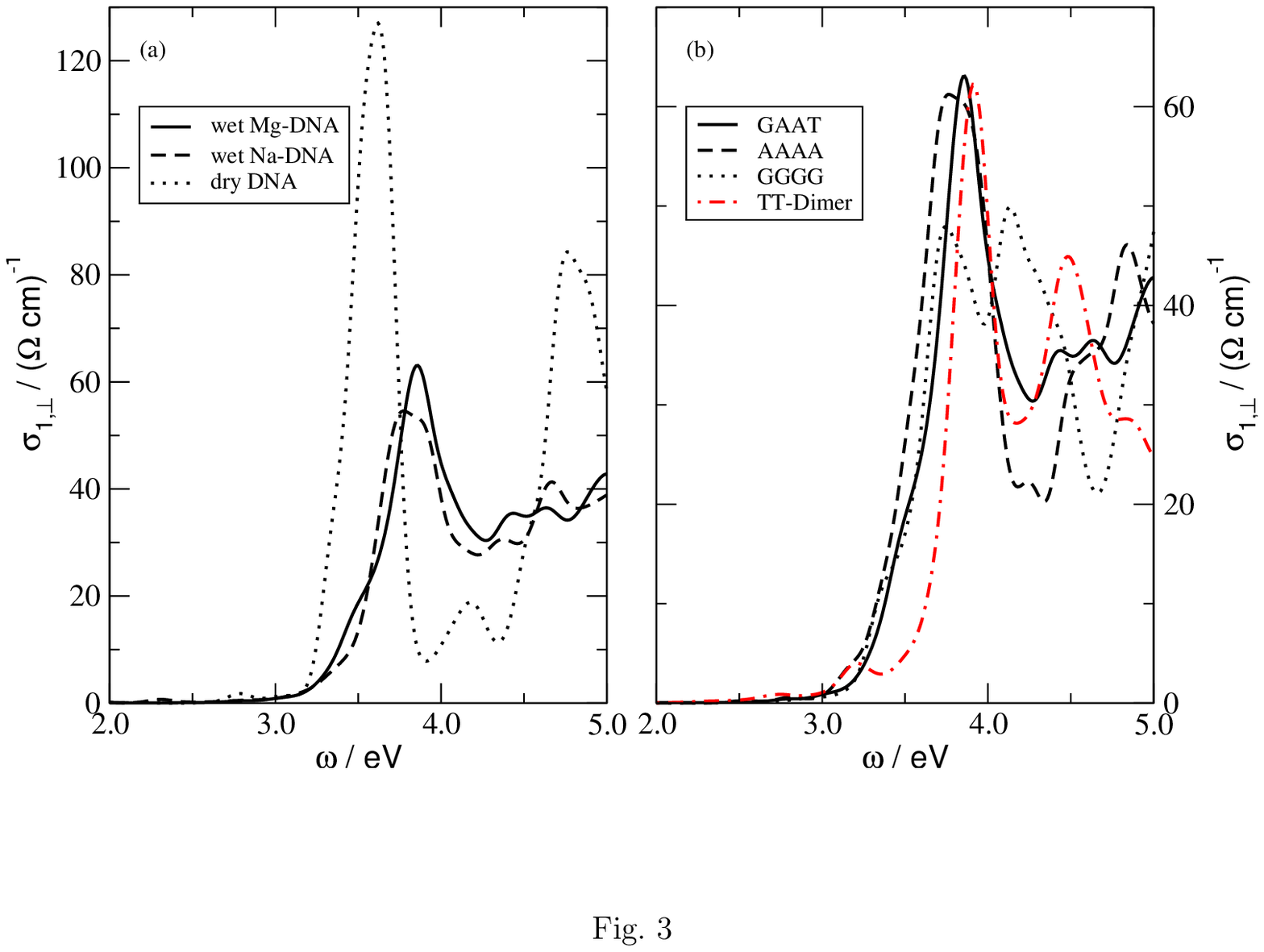}
  }
\end{center}
\caption{
  Robustness of the $\pi$-$\pi^{*}$ transition regarding
  environment and DNA sequence. Panel (a) shows the effects of the
  surrounding water molecules and counterions on the optical conductivity of a
  5'-GAAT-3' sequence. The results for different B-DNA sequences in a wet
  environment containing Mg counterions are compared in panel (b). The
  spectra for the wet 5'-GAAT-3' (5'-AAAA-3', 5'-GGGG-3', and TT-dimer)
  sequences have been averaged over the results of 9 (5) structure
  snap shots. The theoretical line spectra are broadened with Gaussian
  functions of width $0.1$ eV. 
}
\label{Fig_seq}
\end{figure}

In order to identify the energetically important states near
the Fermi energy, we projected the density of states (DOS) on the atomic
$p_{z}$ orbitals of the base pairs and on the water molecules. We
also calculated the projected DOS of the counterions, phosphates, and the
sugars. However, we only found contributions of the $p_{z}$ orbitals and the
water molecules near the Fermi level. A typical projected DOS of structure
snapshot of a 5'-GAAT-3' sequence with Mg counterions is shown in
Fig.~\ref{Fig_pdos}. As seen in panel (a), there is a rather large
$\pi$-$\pi^{*}$ gap of about 2.7 eV. However, if we project the DOS on
a single base we would find an intrabase $\pi$-$\pi^{*}$ gap of about
3.7 eV. It turns out that only such intrabase transitions have large dipole
matrix elements. Therefore, we observe optical gaps of the same size (compare
Figs.~\ref{Fig_pol} and \ref{Fig_seq}). 

Although a rather big $\pi$-$\pi^{*}$ gap is observed the actual gap
between the highest occupied molecular orbital (HOMO) and the lowest
unoccupied molecular orbital (LUMO) of about 150 meV is very small [combine
panels (a) and (b) of Fig.~\ref{Fig_pol}]. Thus, electrons could be excited
from water states below the Fermi energy into unoccupied $p_{z}$ orbitals due
to thermal fluctuations. Notice that the water states
just below the Fermi energy belong to water molecules near the ends of the
sequence [see panel (b) of Fig.~\ref{Fig_pdos}]. Therefore, a small gap
between HOMO and LUMO only appears if water molecules could enter the DNA
structure on damage sites like ends, breaks, or nicks to contact the DNA
bases. 

The results of the projected DOS suggest the possibility of a thermally
activated doping of the DNA by water states which should also lead to an
electronic contribution to the low-frequency absorption. 
We indeed observe low-energy features in the
optical conductivity [panel (a) Fig.~\ref{Fig_pol}]. 
By studying the (fictitious) temperature dependence of
these features within SIESTA, we confirm that they are due to transitions
between thermally occupied $\pi^*$ states of the DNA bases. Furthermore, as
shown in panel (b), we also 
observe the pronounced $\pi$-$\pi^{*}$ transition at
$3.8$~eV. Whereas the polarization dependence of the low-energy absorption is
rather small [see panel (a)] the in-plane $\pi$-$\pi^{*}$ transition
is much more visible if the electric field is perpendicular to the DNA strand
direction. A less pronounced anisotropy of the optical conductivity has also
been found in Ref.~\onlinecite{Gervasio}.

To take into account the dynamical character of the environment we average
the results for the optical conductivity over up to 9 structure snapshots from
the MD simulations. As discussed above, the low-energy features of the optical
conductivity result from a doping of the DNA by water states. Therefore, they
are strongly affected by environment fluctuations and we only obtain some
smeared intensity in the low-energy range of the averaged conductivity. In
contrast, the intrinsic $\pi$-$\pi^{*}$ gap is much less affected
by dynamical fluctuations leading to well defined structures in that
range. 

Due to the intrinsic character of the $\pi$-$\pi^{*}$
transition the averaged optical conductivity in the range $2-5$
eV is barely affected by different counterions [see panel (a) of
Fig.~\ref{Fig_seq}]. To show that the main structure in the spectra around
$3.7$ eV really has to be interpreted as the intrabase 
$\pi$-$\pi^{*}$ transition we also plot the optical
conductivity of dry protonated B-DNA in panel (a) of
Fig.~\ref{Fig_seq}. Although the spectra of wet and dry B-DNA considerably
differ in the intensities, the peak positions agree quite well.
Furthermore, as seen from panel (b) of Fig.~\ref{Fig_seq}, peak
position and shape of the $\pi$-$\pi^{*}$ transition depend
only weakly on sequence whereas remarkable differences are
observed in the range $4-5$ eV.  

\begin{figure}
\begin{center}
  \scalebox{0.49}{
    \includegraphics*[60,420][550,710]{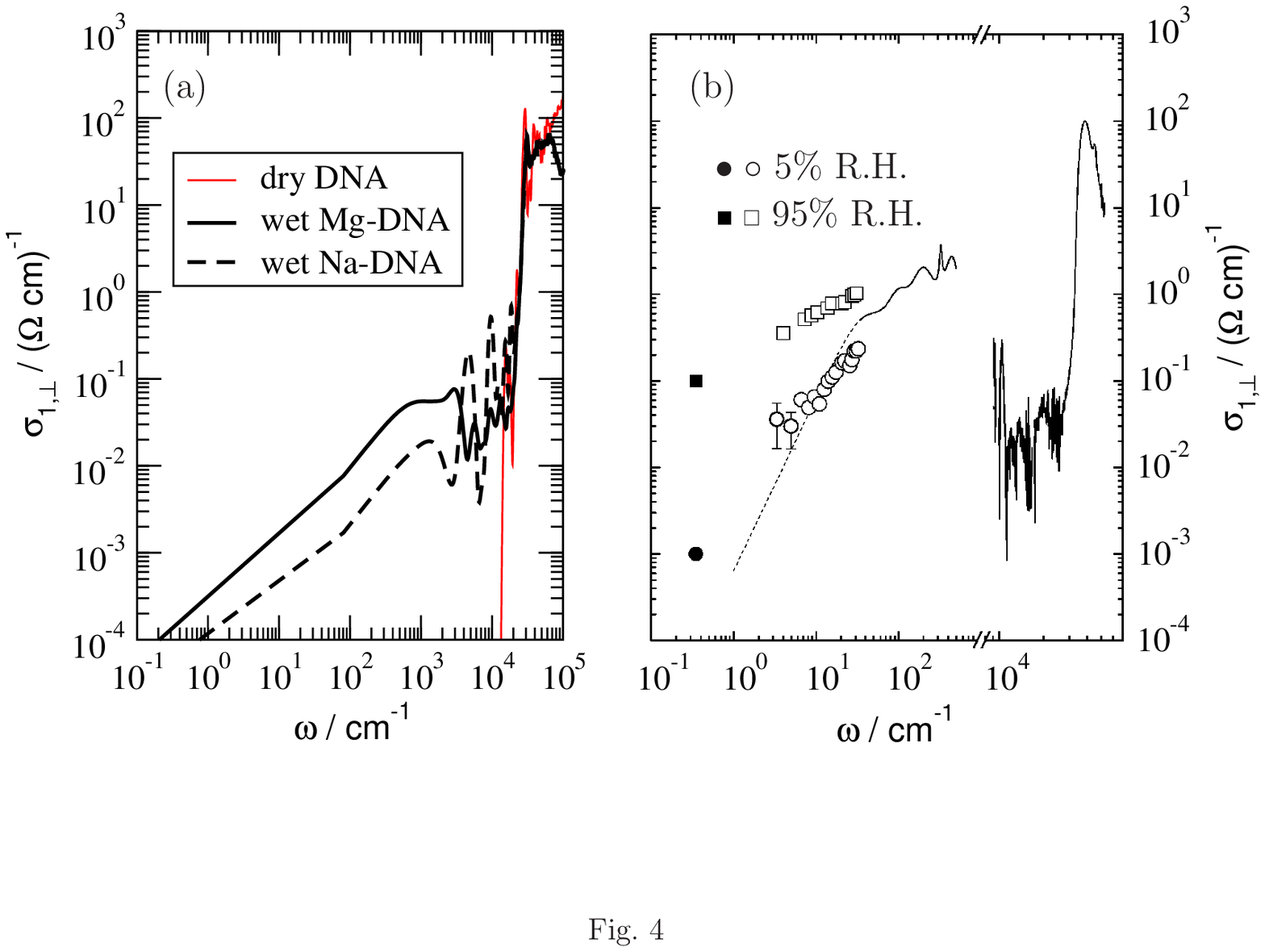}
  }
\end{center}
\caption{
  Panel (a) is a log-log plot of the optical conductivity from
  Fig.~\ref{Fig_seq}, panel (a). Panel (b) shows experimental data for DNA in
  a 5\% and 95\% relative humidity environment taken from
  Ref.~\onlinecite{Helgren}. 
}
\label{Fig_exp}
\end{figure}

In Fig.~\ref{Fig_exp} we compare our results with recent optical experiments
\cite{Helgren} and find a nice agreement for the range of the 
$\pi$-$\pi^{*}$ transition above $10000\,cm^{-1}$. However,
we find pronounced polarization dependence of the conductivity (see
Fig.~\ref{Fig_pol}) whereas the experimental spectra are almost isotropic. 
The latter can be explained by substantial variations in the
axis of the macroscopically orientated DNA duplex \cite{Helgren}.

The spectra below $500\,cm^{-1}$ are also affected by
vibrational modes of the double helix structure and the 
surrounding water molecules, which are not included in our DFT
calculations. In particular, the optical conductivity below $100\,cm^{-1}$ has
been interpreted in Ref.~\onlinecite{Briman} as a pure water dipole
relaxation. However, our DFT calculations clearly show an electronic
conductivity even at very low frequencies (see Fig.~\ref{Fig_exp}). As already
discussed above, the low-frequency conductivity results from a doping of the
DNA by water molecules near the bases. Therefore, the theoretically observed
conductivity at low frequencies, which is approximately an order of
magnitude smaller than the experimental values, would be further increased if
more waters could enter the DNA structure on additional damage sites like ends,
breaks, or nicks.  

A striking aspect of the work of Briman {\it et al.} is the near identity of
the microwave conductivity for single stranded and double stranded
DNA \cite{Briman}.  This, as well as the order of magnitude discrepancy  
between our low frequency calculations and their data can possibly be resolved
as follows: (i) We have order one water per tetramer associated with end
bases; if this is upped to one level per base we gain a factor of eight in
absorption intensity, close to that needed to resolve the discrepancy with the
Briman {\it et al.} data. This is plausible for single stranded DNA
adsorbed to the sapphire substrate of Ref.~\onlinecite{Briman}. (ii) To get
the same low frequency conductivity contribution for double stranded DNA
requires that the DNA flatten and unwind (the helix) as has
been observed elsewhere for DNA on surfaces \cite{flattening}.  In the unwound
configuration the extra space between bases allows for entry by water
molecules.   Obviously some of the microwave absorption will be caused by
water motion alone as suggested by Ref.~\onlinecite{Briman}; our purpose here
is to note that electronic absorption affiliated with defect states can be
comparable. 

A grand technological challenge of our times is to develop
a rapid scheme for sequence differentiation in DNA, for example to be able
to differentiate between alleles of a given gene.
Current technologies, such as those of real-time or quantitative PCR,
rely on fluorescent dyes and the FRET (Fluorescent Resonance Energy Transfer)
method. This requires developing gene specific labeling dyes, or
donors and acceptors, which will lead to fluorescent signals reflecting the
amount of a given double-stranded DNA sequence in the system. However, the 
intrinsic electronic \cite{Zwolak} and
optical properties of DNA also vary with sequence.
So, a natural question is: would it be possible to
develop label-free optical fingerprinting of DNA that would allow one
to infer different alleles directly from optical experiments without the
use of dyes? 

Optical fingerprinting would require two things to happen. First, one would
need optical signatures predominantly from a given gene and one would need
signals from different alleles to be sufficiently different. The former
may be achieved by gene-chip and by surface enhancement techniques. We
can address here the latter question, how different are the spectra from
different sequences and consequently how easy would it be to establish
such a difference experimentally. Clearly the spectra from different
sequences strongly overlap and this makes it harder to distinguish them. 
Having large copies of the gene generated through polymerase chain reaction 
will lead to signal averaging and
the time averaged signal can be distinguished if there is sufficient accuracy.
For example, we have about 30\% statistical variance for five
configurations used in averaged conductivity calculations of tetramers
while the means diverge at the $\approx 5$\% level.  Hence, we could expect
single point frequency sampling to discern the sequences here with 
a configuration increase to 180.  This number could likely be reduced with
judicious  use of multipoint frequency sampling.  If a simple definitive
fingerprint is required without full sequence information, 
as might be appropriate in applications in forensics or pathogen detection, 
numbers in the ballpark of 100-1000 clones of a given sequence 
could be sufficient.  This compares with the $10^5-10^6$ levels of clones
necessary for current state of the art DNA sequencing
technology\cite{Venter98}. Quantitative theoretical studies will be an
important aid in developing such technologies.

\bigskip
We thank G. Gr\"{u}ner for the permission to use the plot of the experimental
data from Ref.~\onlinecite{Helgren}.
This work was supported by NSF grant PHY 0120999 (the Center for Biophotonics
Science and Technology) and DMR-0240918, 
by the US Department of Energy, Division of Materials
Research, Office of Basic Energy Science, and by DFG grant HU 993/1-1. DLC is
grateful for the support of the Center for Theoretical Biological Physics 
through NSF grants PHY0216576 and PHY0225630, and the J.S. Guggenheim Memorial
Foundation.


\end{document}